\documentclass[intlimits,twoside,a4paper]{article}

\usepackage[cp1251]{inputenc}
\usepackage[eqsecnum]{cmpj3}

\issue{2021}{24}{3}{33606}
\doinumber{10.5488/CMP.24.33606}
\title[Electrostatic screening in charged matrices]%
{Screening of ion-ion correlations in electrolyte solutions adsorbed in charged disordered matrices: Application of replica Ornstein-Zernike equations%
\thanks{Dedicated to Professor Yura V. Kalyuzhnyi on the occasion of his 70th birthday.}}
\author[T. Mlakar, B. Hribar-Lee]{T. Mlakar\orcid{0000-0003-1121-7785},
        B. Hribar-Lee\orcid{0000-0002-9029-588X}\thanks{Corresponding author: \email{barbara.hribar@fkkt.uni-lj.si}.}}
\address{
 University of Ljubljana, Faculty of Chemistry and Chemical Technology, Ve{\v c}na pot 113, 1000 Ljubljana, Slovenia 
}

\Keywords{electrolyte screening, charged matrix, templated matrix, replica Ornstein-Zernike equations}

\date{Received June 08, 2021, in final form July 29, 2021}

\begin{document}

\maketitle

\begin{abstract}
The replica Ornstein-Zernike equations for an electrolyte adsorbed in a charged, disordered matrix were applied to a model, where both subsystems consisted of points carrying a single (positive or negative) charge. While the system as a whole was electroneutral, each of the subsytems had a net charge. The results of this study are compared with the ones previously obtained, where the interactions in such a system were considered to be the same as in the case of electroneutral subsystems.
%
%
\printkeywords
%
\end{abstract}

\section{Introduction}
Disordered porous materials that serve as adsorbents are of considerable interest, not just for basic research, but for various applications, such as: separation sciences, medicine, and catalysis~\cite{Jardat2017,Dolce2018,Bacle2016,LeGoas2020}. Systems of this kind can be viewed as partly-quenched, where some degrees of freedom are quenched (frozen), while others are annealed (freely moving). Accordingly, we can use statistical-mechanical theories for the description of such systems~\cite{Given1994,Madden1992,Trokhymchuk1996}. One of the approaches that has been developed for studying partly quenched systems containing charges is the replica Ornstein-Zernike (ROZ) theory~\cite{Hribar1997,Hribar2011}. In correspondence with this theory, the correlation functions are divided into the ``connecting'' part, representing the interactions between ions within the same replica, and into the ``blocking'' part, describing the interaction mediated by matrix particles~\cite{Given1994,Madden1992,Hribar2011}. Equations can be used within the integral equation theory approximations to obtain structural, as well as thermodynamic properties of partly quenched systems, that are in good agreement with computer simulation results.

The ROZ equations were also extended to describe partly quenched systems where a templated matrix was used. Templated particles were present during the matrix equilibration process but were removed after the quench~\cite{Zhang2000,Zhang2000b}. These equations have also been used to study partly quenched systems containing charges~\cite{Dominguez2003,Luksic2007}; the matrix was prepared by equilibrating an electrolyte solution, and after the quench, only cations remained frozen (the matrix itself was carrying a net positive charge), while anions became part of the annealed fluid (which was also carrying a net charge, although the system as a whole was electroneutral). In studies of this kind, it has been assumed that the long-range parts of the correlation functions remain the same as in the case of non-templated, electroneutral subsystems.

In the present work we use a rigorous derivation of the renormalization scheme for partly quenched systems containing charges, where quenched and annealed components are charged. This paper is organised as follows: after the short introduction, we describe the model under consideration. We continue by presenting the theoretical procedure, and showing some numerical results. Conclusions are given in the end.


\section{The model description}

The system under study consists of two subsystems, the first one is called matrix, and the second is an annealed ionic fluid. The notation used in this paper is similar to the used before: the superscripts $0$, $0'$, and $1$ correspond to the matrix, the template, and the annealed fluid species, respectively~\cite{Hribar2011,Luksic2007,Dominguez2003}.

In our model, the matrix is obtained by equilibrating a $+1$:$-1$ primitive model electrolyte with the number density $\rho _+ ^0 = \rho _- ^0 = \rho ^0$ at temperature $T_0$. After equilibration, the cations remain frozen, while the anions (that served as a template) become annealed. The adsorbing electrolyte is also a $+1$:$-1$ primitive model electrolyte. Note that the number density of annealed cations is lower than the number density of annealed anions since the adsorbing electrolyte also contains the annealed ions of the matrix: $\rho _+ ^1  + \rho _+ ^0 = \rho_- ^1 + \rho_- ^{0'}$; the system as a whole is electroneutral. All the ions are singly charged. The system is studied at temperature $T_1$, which in general can differ from the temperature $T_0$. The ratio $\varepsilon_0 T_0/\varepsilon_1 T_1=Q$ is called quenching parameter. Here $\varepsilon _i$ is the dielectric constant of the solution at corresponding temperature.

The ions in this article are modelled as points carrying a single (positive or negative), charge so the interactions between them can be written as:

\begin{align}
\Phi^{00}_{ij}&=-\frac{z_i^0 z_j^0 e_0^2}{4\piup\varepsilon \varepsilon_0 k_{\rm{B}}T_0r} = -\frac{1}{Q}\frac{ z_i^0 z_j^0 L_b}{r} ,\nonumber \\
 \Phi^{10}_{ij}&=-\frac{z_i^1 z_j^0 e_0^2}{4\piup\varepsilon \varepsilon_1 k_{\rm{B}}T_1r} = -\frac{z_i^1 z_j^0 L_b}{r}, \nonumber \\
 \Phi^{11}_{ij}&=-\frac{z_i^1 z_j^1 e_0^2}{4\piup\varepsilon \varepsilon_1 k_{\rm{B}}T_1r} = -\frac{z_i^1 z_j^1 L_b}{r},  
\end{align}

\noindent $e_0$ denotes the elementary charge, $\varepsilon$ is the permittivity of vacuum, and $k_{\rm{B}}$ is Boltzmann constant. $z_i ^m$ are valencies of the ions. $L_b$ is the so-called Bjerrum length, defined as $L_b = e_0^2 / 4\piup\varepsilon \varepsilon_0 k_{\rm{B}}T_1$.

Note that there are no interactions between ions in different replicas ($\Phi^{12}=0$), and there are no interactions between annealed ions and the template ($\Phi^{10'}=0$).

\section{Theoretical procedure}

In the case where the structure of the matrix is obtained by being treated as a template, the set of ROZ equations can be written as~\cite{Zhang2000,Zhang2000b}:

\begin{eqnarray}
    h^{10} = c^{10} + \rho^0 c^{10}\otimes h^{00} + \rho^{0'} c^{10'}\otimes h^{0'0}+ \rho^1 c^{11}\otimes h^{10} -    \rho^1 c^{12}\otimes h^{10} , \nonumber\\
    h^{10'} = c^{10'} + \rho^0 c^{10}\otimes h^{00'} + \rho^{0'} c^{10'}\otimes h^{0'0'}+ \rho^1 c^{11}\otimes h^{10'} -    \rho^1 c^{12}\otimes h^{10'} , \nonumber\\
    h^{11} = c^{11} + \rho^0 c^{10}\otimes h^{01} + \rho^{0'} c^{10'}\otimes h^{0'1}+ \rho^1 c^{11}\otimes h^{11} -    \rho^1 c^{12}\otimes h^{21} ,\nonumber\\
    h^{12} = c^{12} + \rho^0 c^{10}\otimes h^{01} + \rho^{0'} c^{10'}\otimes h^{0'1}+ \rho^1 c^{11}\otimes h^{12} + \rho^1 c^{12}\otimes h^{11}- 2 \rho^1 c^{12}\otimes h^{21},
\label{eq:oz}
\end{eqnarray}

\noindent where the symbol $\otimes$ denotes convolution, $c^{mn}$ the direct correlation function and $h^{mn}$ the total correlation function. They are $2\times 2$ matrices for electrolyte solutions, that  contain elements $++$, $+-$, $-+$, and $--$. $\rho^m$ is a $2\times 2$ diagonal matrix containing the number density of cations and anions. Since there are no short-range interactions between model ions for our system, both kinds of correlation functions only consist of a long range part:

\begin{eqnarray}
c^{mn}_{ij}=\Phi^{mn}_{ij},\nonumber \\
h^{mn}_{ij}=q^{mn}_{ij}.
\end{eqnarray}

From equation~(\ref{eq:oz}) we obtain:

\begin{align}
    q^{10}_{i0}-\Phi^{10}_{i0}&=\Phi^{10}_{i0}\otimes\rho^0q^{00}_{00}+\Phi^{11}_{ii}\otimes\rho^1_i q^{10}_{i0}+\Phi^{11}_{ij}\otimes\rho^1_j  q^{10}_{j0}, \nonumber \\
    q^{10^{'}}_{i0^{'}}&=\Phi^{10}_{i0}\otimes\rho^0q^{00^{'}}_{00^{'}}+\Phi^{11}_{ii}\otimes\rho^1_i q^{10^{'}}_{i0^{'}}+\Phi^{11}_{ij}\otimes\rho^1_j q^{10^{'}}_{j0^{'}},\nonumber \\
    q^{11}_{ij}-\Phi^{11}_{ij}&=\Phi^{10}_{i0}\otimes\rho^0q^{01}_{0j}+\Phi^{11}_{ii}\otimes\rho^1_i q^{11}_{ij}+\Phi^{11}_{ij}\otimes\rho^1_j q^{11}_{jj},\nonumber \\
    q^{12}_{ij} &= \rho^0 \Phi^{10}_{i0}\otimes q^{01}_{0j} + \rho^1 _i \Phi^{11}_{ii}\otimes q^{12}_{ij} + \rho^1 _j \Phi^{11}_{ij}\otimes q^{12}_{jj}.
\label{eq:renorm}
\end{align}\\

We proceed by obtaining the long range parts of the total correlation functions. 

\subsection{Fluid-matrix (10) correlations}

We begin with the correlation functions between the fluid and the matrix. Since the fluid consists of two components, cations ($z^1 _+ = +1$), and anions ($z^1 _- = -1$), we can rewrite the first equation of~(\ref{eq:renorm}) into the following matrix form, using the Fourier transform of correlation functions (denoted by $ \phantom{!}\widetilde{}\phantom{1}$):

\begin{equation}
\renewcommand\arraystretch{1.2}
\begin{bmatrix}
(1-\widetilde{\Phi}^{11}_{++}\rho^1_+)&
(-\widetilde{\Phi}^{11}_{+-}\rho^1_-)\\
(-\widetilde{\Phi}^{11}_{-+}\rho^1_+)&
(1-\widetilde{\Phi}^{11}_{--}\rho^1_-)\\
\end{bmatrix}\begin{bmatrix}
\widetilde{q}^{10}_{+0}\\
\widetilde{q}^{10}_{-0}\\
\end{bmatrix}=
\begin{bmatrix}
\widetilde{\Phi}^{10}_{+0}(1+\rho^0\widetilde{q}^{00}_{00})\\
\widetilde{\Phi}^{10}_{-0}(1+\rho^0\widetilde{q}^{00}_{00})\\
\end{bmatrix}.
\end{equation}

\noindent From which $\widetilde{q}^{10}_{ij}$ can be expressed as:

\begin{equation}
\renewcommand\arraystretch{1.2}
\begin{bmatrix}
\widetilde{q}^{10}_{+0}\\
\widetilde{q}^{10}_{-0}\\
\end{bmatrix}=
\begin{bmatrix}
(1-\widetilde{\Phi}^{11}_{++}\rho^1_+)&
(-\widetilde{\Phi}^{11}_{+-}\rho^1_-)\\
(-\widetilde{\Phi}^{11}_{-+}\rho^1_+)&
(1-\widetilde{\Phi}^{11}_{--}\rho^1_-)\\
\end{bmatrix}^{-1}
\begin{bmatrix}
\widetilde{\Phi}^{10}_{+0}(1+\rho^0\widetilde{q}^{00}_{00})\\
\widetilde{\Phi}^{10}_{-0}(1+\rho^0\widetilde{q}^{00}_{00})\\
\end{bmatrix}.
\end{equation}\\

\noindent Taking into account that in our model $\widetilde{\Phi}^{11}_{++} = \widetilde{\Phi}^{11}_{--}= - \widetilde{\Phi}^{11}_{+-}$ and $\widetilde{\Phi}^{10}_{+0}= - \widetilde{\Phi}^{10}_{-0}$, one obtains:

\begin{equation}
\renewcommand\arraystretch{1.2}
\begin{bmatrix}
\widetilde{q}^{10}_{+0}\\
\widetilde{q}^{10}_{-0}\\
\end{bmatrix}=
\frac{1}{1-\widetilde{\Phi}^{11}_{++}(\rho^1_++\rho^1_-)}
\begin{bmatrix}
\widetilde{\Phi}^{10}_{+0}(1+\rho^0\widetilde{q}^{00}_{00})\\
-\widetilde{\Phi}^{10}_{+0}(1+\rho^0\widetilde{q}^{00}_{00})\\
\end{bmatrix}.
\end{equation}
Taking into further account that:

\begin{equation}
\widetilde{q}^{00}_{00}=-\frac{4\piup L_b}{Q(k^2+k_0^2)},
\end{equation}

\noindent where $ k_0 ^2 =\frac{4\piup L_b}{Q} [\rho^0z^{0}z^{0}+\rho^{0'}z^{0'}z^{0'}]$, we get: 

\begin{equation}
\renewcommand\arraystretch{1.2}
\begin{bmatrix}
\widetilde{q}^{10}_{+0}\\
\widetilde{q}^{10}_{-0}\\
\end{bmatrix}=
\frac{(-z^{0}z^{1}_+)\frac{e^2}{4\piup\epsilon\epsilon_1 k_{\rm{B}}T_1}\mathcal{F}(1/r)}{1+z^{1}_+z^{1}_+\frac{e^2}{4\piup\epsilon\epsilon_1 k_{\rm{B}}T_1}\mathcal{F}(1/r)(\rho^1_++\rho^1_-)}
\begin{bmatrix}
1-\frac{4\piup L_b\rho^0}{Q(k^2+k_0^2)}\\
-\left( 1-\frac{4\piup L_b\rho^0}{Q(k^2+k_0^2)}\right) \\
\end{bmatrix},
\end{equation}

\noindent where $\mathcal{F}(1/r)$ denotes the Fourier transform of $(1/r)$ with the well known expression $\mathcal{F}(1/r) = 4\piup / k^2$. In our case $z^{0} = z^{0} _+$, and $ z^{0'} = z^{0'} _- $. The final result for $\widetilde{q}^{10}_{ij}$ is as follows:

\begin{equation}
\renewcommand\arraystretch{1.2}
\begin{bmatrix}
\widetilde{q}^{10}_{+0}\\
\widetilde{q}^{10}_{-0}\\
\end{bmatrix}=
\frac{4\piup z^{0}z^{1}_+L_b(Q(k^2+k_0^2)-4\piup L_b\rho^0)} {(k^2+4\piup z^{1}_+z^{1}_+L_b(\rho^1_++\rho^1_-))(k^2+k_0^2)Q}
\begin{bmatrix}
-1\\
1\\
\end{bmatrix}.
\label{eq:q10k}
\end{equation}

\noindent By splitting the above expression into a sum of three fractions, one can readily invert the equations into $r$ space obtaining:

\begin{equation}
\renewcommand\arraystretch{1.2}
\begin{bmatrix}
{q}^{10}_{+0}\\
{q}^{10}_{-0}\\
\end{bmatrix}=\left( \frac{ac}{4\piup r Q(k_0^2-b^2)}(\re^{-k_0r}-\re^{-br})+\frac{a}{4\piup r}\re^{-br}\right)
\begin{bmatrix}
-1\phantom{.}\\
\phantom{-}1\phantom{.}\\
\end{bmatrix},
\end{equation}

\noindent where $a=4\piup z^{0}z^{1}_+L_b$, $b=\sqrt{4\piup z^{1}_+z^{1}_+L_b(\rho^1_++\rho^1_-)}$ and $c=4\piup L_b\rho^0$. Note that in the case of single-valent ions described in this paper, $z^{0}$ and $z^{1}_i$ are equal to 1. The derivation, however is valid for arbitrary nominal ionic charges.

\begin{figure}[h!]
	\centering
	\includegraphics[scale=0.5, angle=0]{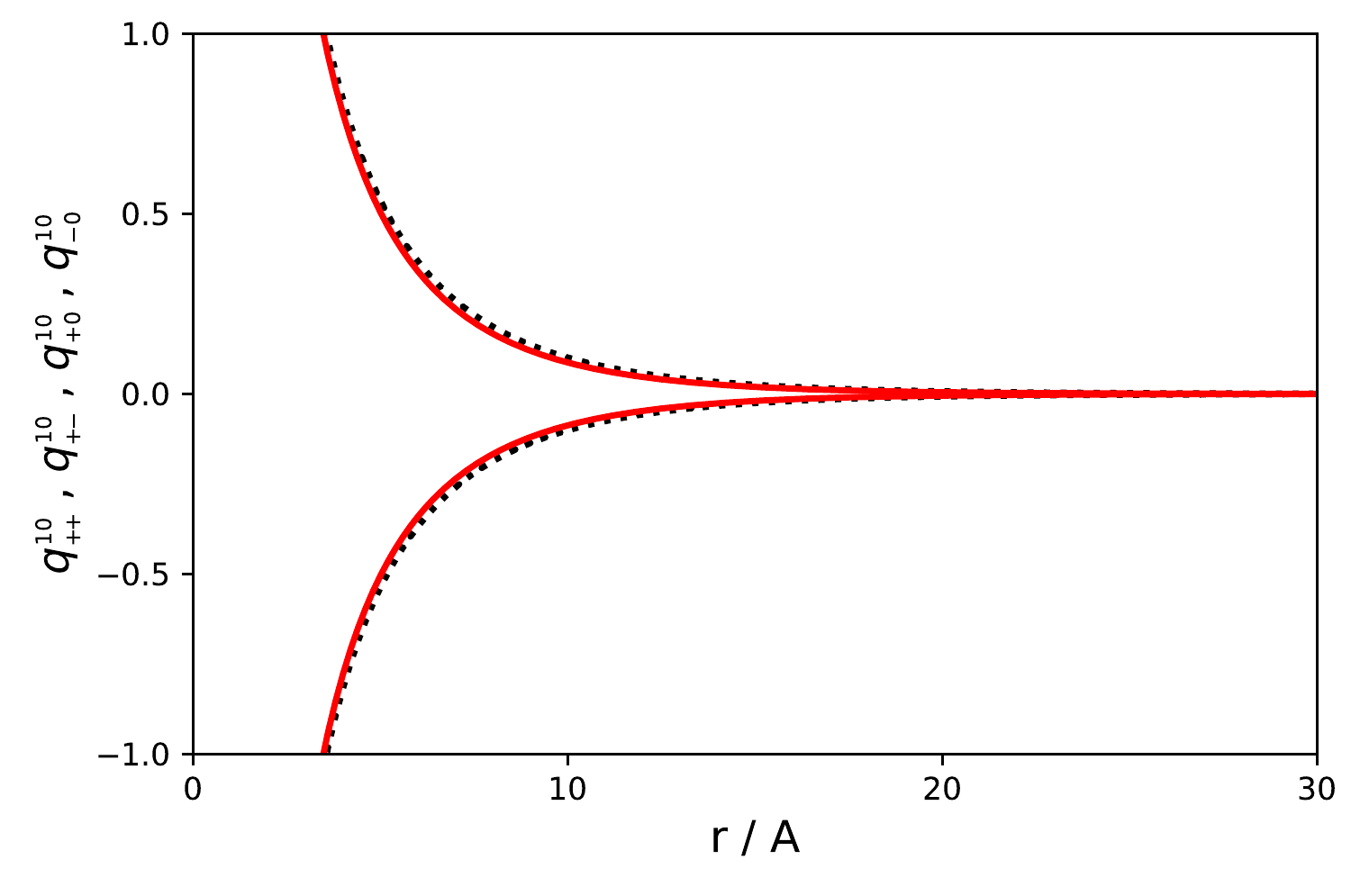}
	\includegraphics[scale=0.5, angle=0]{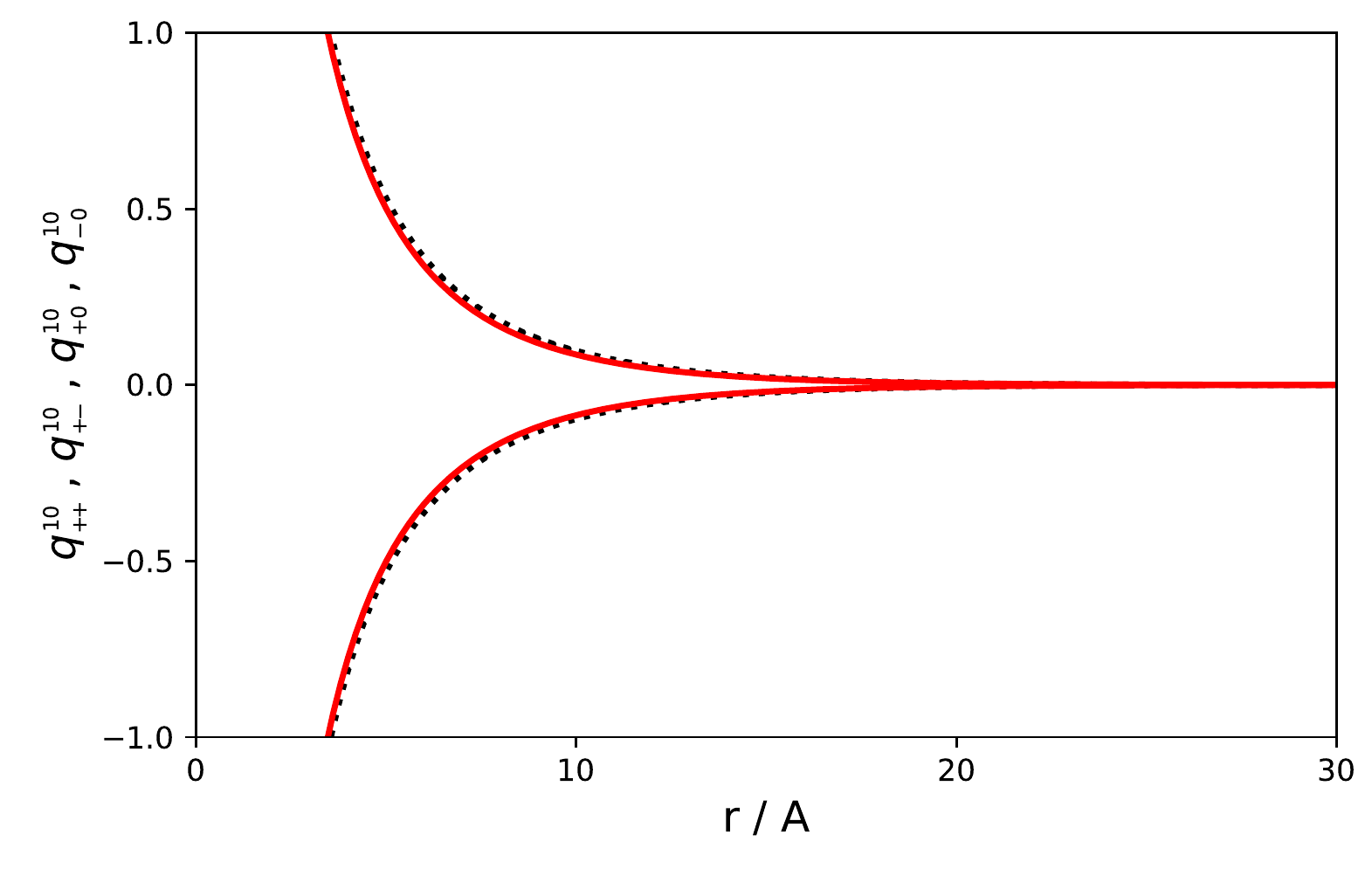}
	\includegraphics[scale=0.5, angle=0]{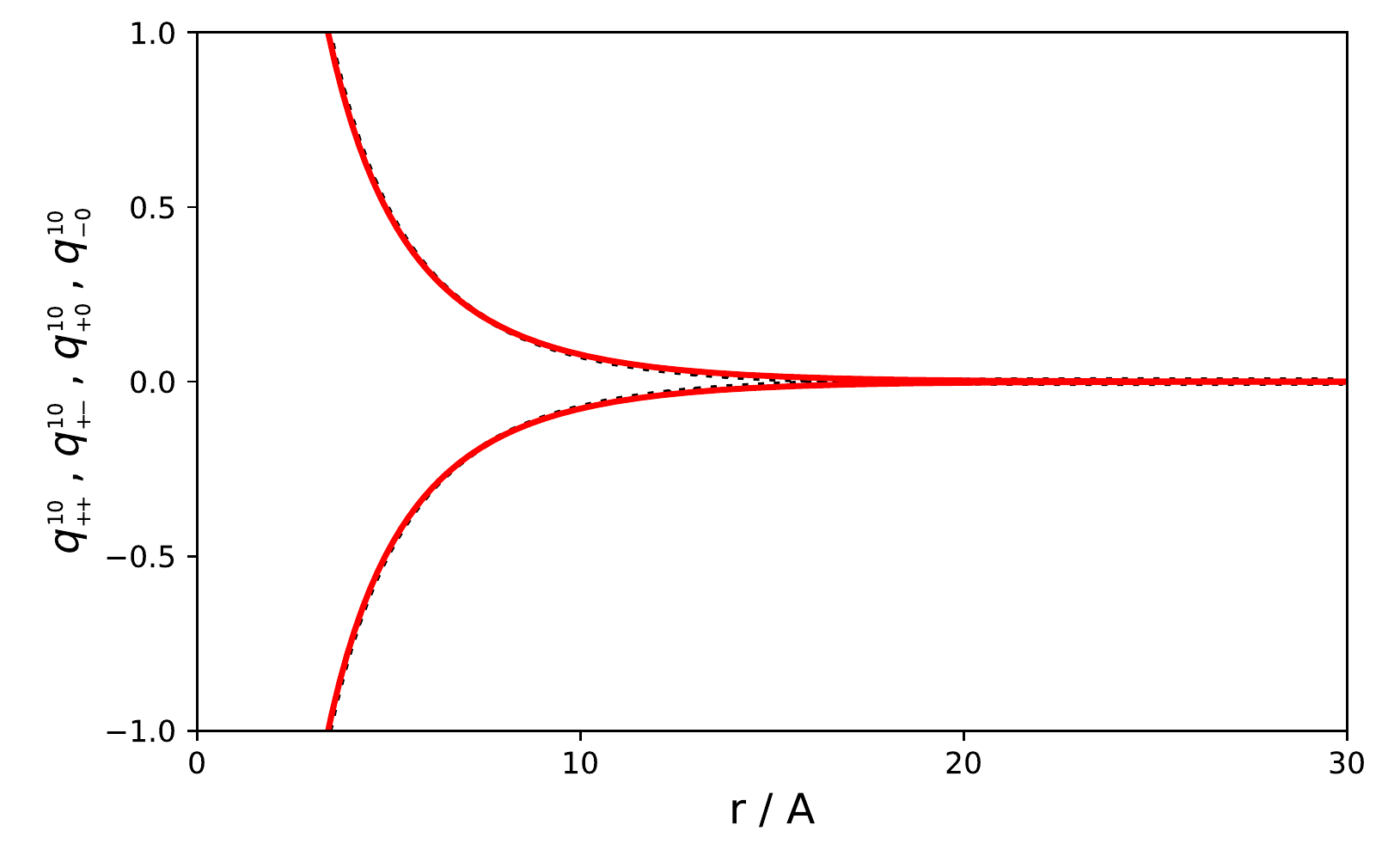}
	\caption{(Colour online) ${q}^{10}_{ij}$ functions for $Q=1.2, L_b=7.14$~\AA, $c_0=0.425$~M ($\rho^{0'}=\rho^0$). $c_1=6.8325 \cdot 10^{-5}$~M (top), $3.187 \cdot 10^{-3}$~M (middle), and $c_1=3.0 \cdot 10^{-2}$~M (bottom). ${q}^{10}_{i0}$ denote the functions obtained in this work, while ${q}^{10}_{ij}$ denote the functions obtained for electroneutral subsystems.}
	\label{Fig:case1}
\end{figure}

\subsection{Fluid-fluid (11) correlations}

Similarly to the case of fluid-matrix correlation functions, we begin by rewriting the third equation of~(\ref{eq:renorm}) for different components of the annealed fluid, which in matrix form reads:

\begin{eqnarray}
\renewcommand\arraystretch{1.2}
&\begin{bmatrix}
(1-\widetilde{\Phi}^{11}_{++}\rho^1_+)&
0\\
0&
(1-\widetilde{\Phi}^{11}_{--}\rho^1_-)\\
\end{bmatrix}
\renewcommand\arraystretch{1.2}
\begin{bmatrix}
\widetilde{q}^{11}_{++}&
\widetilde{q}^{11}_{+-}\\
\widetilde{q}^{11}_{-+}&
\widetilde{q}^{11}_{--}\\
\end{bmatrix}-\begin{bmatrix}
\widetilde{\Phi}^{11}_{++}\rho^1_+&
\widetilde{\Phi}^{11}_{+-}\rho^1_-\\
\widetilde{\Phi}^{11}_{-+}\rho^1_+&
\widetilde{\Phi}^{11}_{--}\rho^1_-\\
\end{bmatrix}\begin{bmatrix}
\widetilde{q}^{11}_{++}&
0\\
0&
\widetilde{q}^{11}_{--}\\
\end{bmatrix}\nonumber \\
& =\begin{bmatrix}
\widetilde{\Phi}^{11}_{++}+\widetilde{\Phi}^{10}_{+0}\rho^0\widetilde{q}^{01}_{0+}&
\widetilde{\Phi}^{11}_{+-}+\widetilde{\Phi}^{10}_{+0}\rho^0\widetilde{q}^{01}_{0-}\\
\widetilde{\Phi}^{11}_{-+}+\widetilde{\Phi}^{10}_{-0}\rho^0\widetilde{q}^{01}_{0+}&
\widetilde{\Phi}^{11}_{--}+\widetilde{\Phi}^{10}_{-0}\rho^0\widetilde{q}^{01}_{0-}\\
\end{bmatrix}.
\label{eq:q11_all}
\end{eqnarray}

Note that due to asymmetry of the system, the matrix form cannot be used directly to obtain the expression for  $\widetilde{q}^{11}_{ij}$. Therefore, we had to develop each expression separately. The expressions for  $\widetilde{q}^{11}_{ij}$ obtained from (\ref{eq:q11_all}) are:

\begin{align}
    \widetilde{q}^{11}_{++}&=\frac{\widetilde{\Phi}^{11}_{++}+\widetilde{\Phi}^{10}_{+0}\rho^0\widetilde{q}^{01}_{0+}}{(1-2\widetilde{\Phi}^{11}_{++}\rho^1_+)}, \nonumber \\
    \widetilde{q}^{11}_{--}&=\frac{\widetilde{\Phi}^{11}_{++}-\widetilde{\Phi}^{10}_{+0}\rho^0\widetilde{q}^{01}_{0-}}{(1-2\widetilde{\Phi}^{11}_{++}\rho^1_-)}, \nonumber \\
     \widetilde{q}^{11}_{+-}&=\frac{-\widetilde{\Phi}^{11}_{++}(1+\rho^1_- \widetilde{q}^{11}_{--})+\widetilde{\Phi}^{10}_{+0}\rho^0\widetilde{q}^{01}_{0-}}{(1-\widetilde{\Phi}^{11}_{++}\rho^1_+)}.
\end{align}

Here, we again took into account that $\widetilde{\Phi}^{11}_{++} = \widetilde{\Phi}^{11}_{--}= - \widetilde{\Phi}^{11}_{+-}$, and $\widetilde{\Phi}^{10}_{+0}= - \widetilde{\Phi}^{10}_{-0}$ for our model.

Since the expression for $\widetilde{q}^{11}_{+-}$ requires $\widetilde{q}^{11}_{++}$, and $\widetilde{q}^{11}_{--}$, we first obtained the solutions for the latter. By inserting the expressions for $\widetilde{q}^{10}_{i0}$ obtained above [equation~(\ref{eq:q10k})], one can show:

\begin{align}
    \widetilde{q}^{11}_{++}&=\frac{-a \frac{z^{1}_+}{z^{0}}+\rho^0\frac{a^{2}(Q(k^2+k_0^2)-c)} {(k^2+b^{2})(k^2+k_0^2)Q}}{k^2+8\piup z^{1}_+z^{1}_+L_b\rho^1_+}, \nonumber \\
     \widetilde{q}^{11}_{--}&=\frac{-a \frac{z^{1}_+}{z^{0}}+\rho^0\frac{a^{2}(Q(k^2+k_0^2)-c)} {(k^2+b^{2})(k^2+k_0^2)Q}}{k^2+8\piup z^{1}_+z^{1}_+L_b\rho^1_-}.
\end{align}

Now, by introducing two more constants, $b_+=\sqrt{8\piup z^{1}_+z^{1}_+L_b\rho^1_+}$, and $b_-=\sqrt{8\piup z^{1}_+z^{1}_+L_b\rho^1_-}$, the $q^{11}_{+-}$ in $k$-space, $\widetilde{q}^{11}_{+-}$ can be written as:

\begin{equation}
    \widetilde{q}^{11}_{+-}=\frac{a \frac{z^{1}_+}{z^{0}}\left(1+\frac{-\frac{\rho^1_-z^{1}_+a}{z^{0}}+\frac{a^{2}\rho^0\rho^1_-(Q(k^2+k_0^2)-c)} {(k^2+b^{2})(k^2+k_0^2)Q}}{k^2+b_-^{2}}\right)-\frac{a^{2}\rho^0(Q(k^2+k_0^2)-c)} {(k^2+b^{2})(k^2+k_0^2)Q}}{k^2+b_+^{2}/2}.
\end{equation}

Similarly to the case of fluid-matrix functions, we proceed by inverting the equations into $r$ space. Introducing two more constants, $\alpha=az^{1}_+ / z^{0}$ and $\beta=a^{2}\rho^0$, we get the final expressions for our desired quantities:

\begin{align}
  {q}^{11}_{++}&=  \frac{\re^{-br}}{4\piup r}\left(\frac{\beta}{b^{2}_+-b^{2}}-\frac{\beta c}{Q}\frac{1}{(b^{2}-k_0^{2})(b^{2}-b^{2}_+)}  \right)+\frac{\re^{-k_0r}}{4\piup r}\frac{\beta c/Q}{(b^{2}-k_0^{2})(k_0^{2}-b^{2}_+)} \nonumber \\&-\frac{\re^{-rb_+}}{4\piup r}\left(\alpha+\frac{\beta}{b^{2}_+-b^{2}}+\frac{\beta c}{Q}\frac{1}{(b^{2}-b^{2}_+)(k_0^{2}-b^{2}_+)}\right),\nonumber \\
  {q}^{11}_{--}&=  \frac{\re^{-br}}{4\piup r}\left(\frac{\beta}{b^{2}_--b^{2}}-\frac{\beta c}{Q}\frac{1}{(b^{2}-k_0^{2})(b^{2}-b^{2}_-)}  \right)+\frac{\re^{-k_0r}}{4\piup r}\frac{\beta c/Q}{(b^{2}-k_0^{2})(k_0^{2}-b^{2}_-)}\nonumber\\&-\frac{\re^{-rb_-}}{4\piup r}\left(\alpha+\frac{\beta}{b^{2}_--b^{2}}+\frac{\beta c}{Q}\frac{1}{(b^{2}-b^{2}_-)(k_0^{2}-b^{2}_-)}\right),\nonumber \\
  q^{11}_{+-}&=\frac{\re^{- rb_+/\sqrt{2}}}{4\piup r}\frac{(\beta(b^{2}_+/2-b_-^{2})+\rho^1_-\alpha^{2}(b^{2}_+/2-b^{2})+\alpha\beta\rho^1_-)(b^{2}_+/2-k_0^{2})Q} {(b^{2}_+/2-b_-^{2})(b^{2}_+/2-b^{2})(b^{2}_+/2-k_0^{2})Q} \nonumber \\&+\frac{\re^{-br}}{4\piup r}\frac{(\beta(b^{2}_--b^{2})-\alpha\beta\rho^1_-)(b^{2}-k_0^{2})Q+\beta c(b^{2}_--b^{2})-\alpha\beta\rho^1_-c} {(b^{2}-b^{2}_+/2)(b^{2}_--b^{2})(b^{2}-k_0^{2})Q} \nonumber \\&+\frac{\re^{- rb_+/\sqrt{2}}}{4\piup r}\left(\frac{\beta c(\alpha\rho^1_-+(b^{2}_+/2-b_-^{2}))}{(b^{2}_+/2-b_-^{2})(b^{2}_+/2-b^{2})(b^{2}_+/2-k_0^{2})Q}+\alpha\right)
\nonumber
\\&+\frac{\re^{- rb_-}}{4\piup r}\frac{(\rho^1_-\alpha^{2}(b^{2}_--b^{2})+\alpha\beta\rho^1_-)(b^{2}_--k_0^{2})Q+\alpha\beta\rho^1_-c} {(b_-^{2}-b^{2}_+/2)(b^{2}_--b^{2})(b^{2}_--k_0^{2})Q}\nonumber \\&+\frac{\re^{-k_0r}}{4\piup r}\frac{\beta c(b^{2}_--k_0^{2})-\alpha\beta\rho^1_-c} {(b^{2}_+/2-k_0^{2})(b^{2}_--k_0^{2})(b^{2}-k_0^{2})Q}.
\end{align}

\section{Numerical results}

\begin{figure}[ht!]
\centering
\includegraphics[scale=0.52, angle=0]{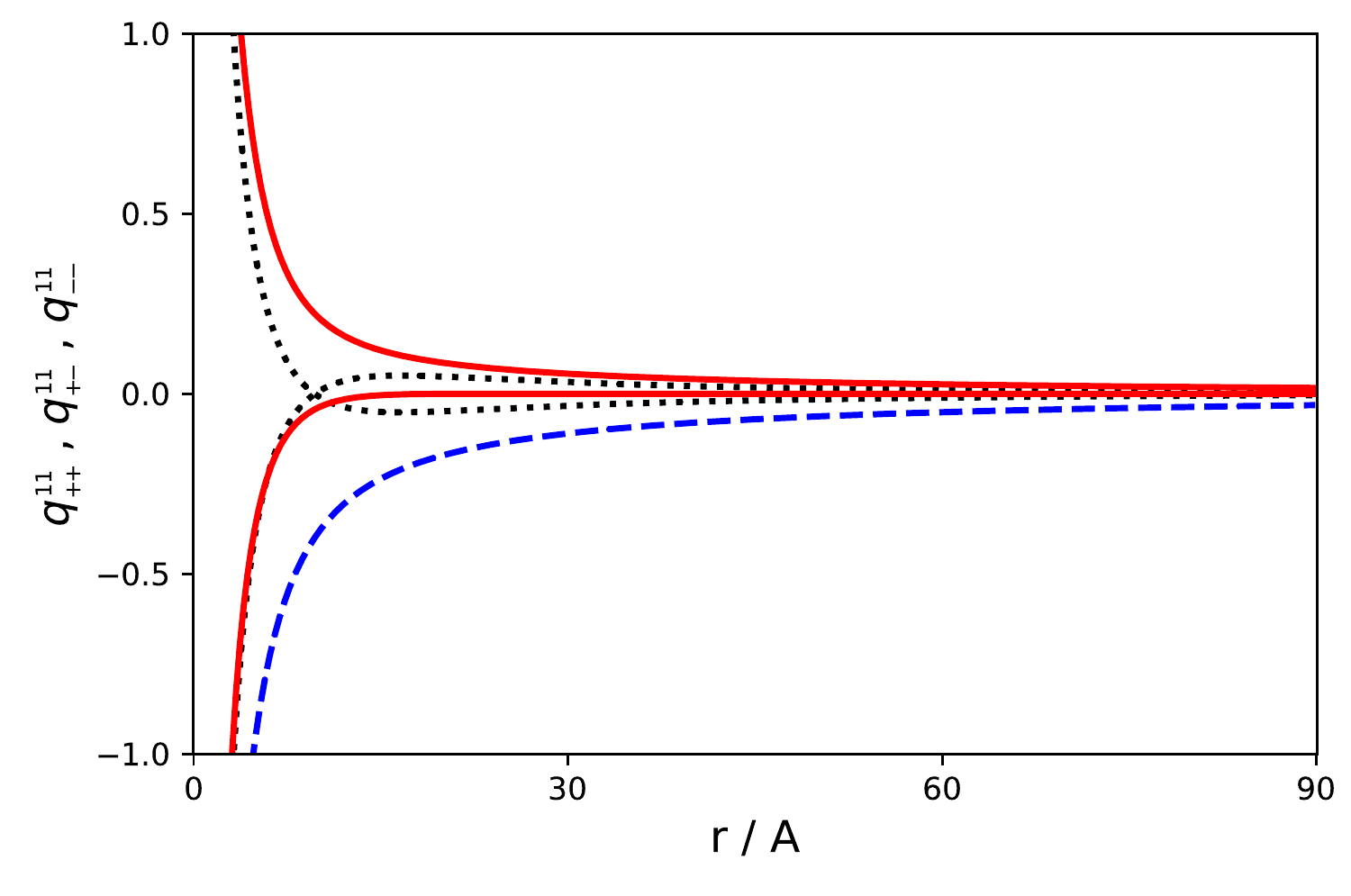}
\includegraphics[scale=0.45, angle=0]{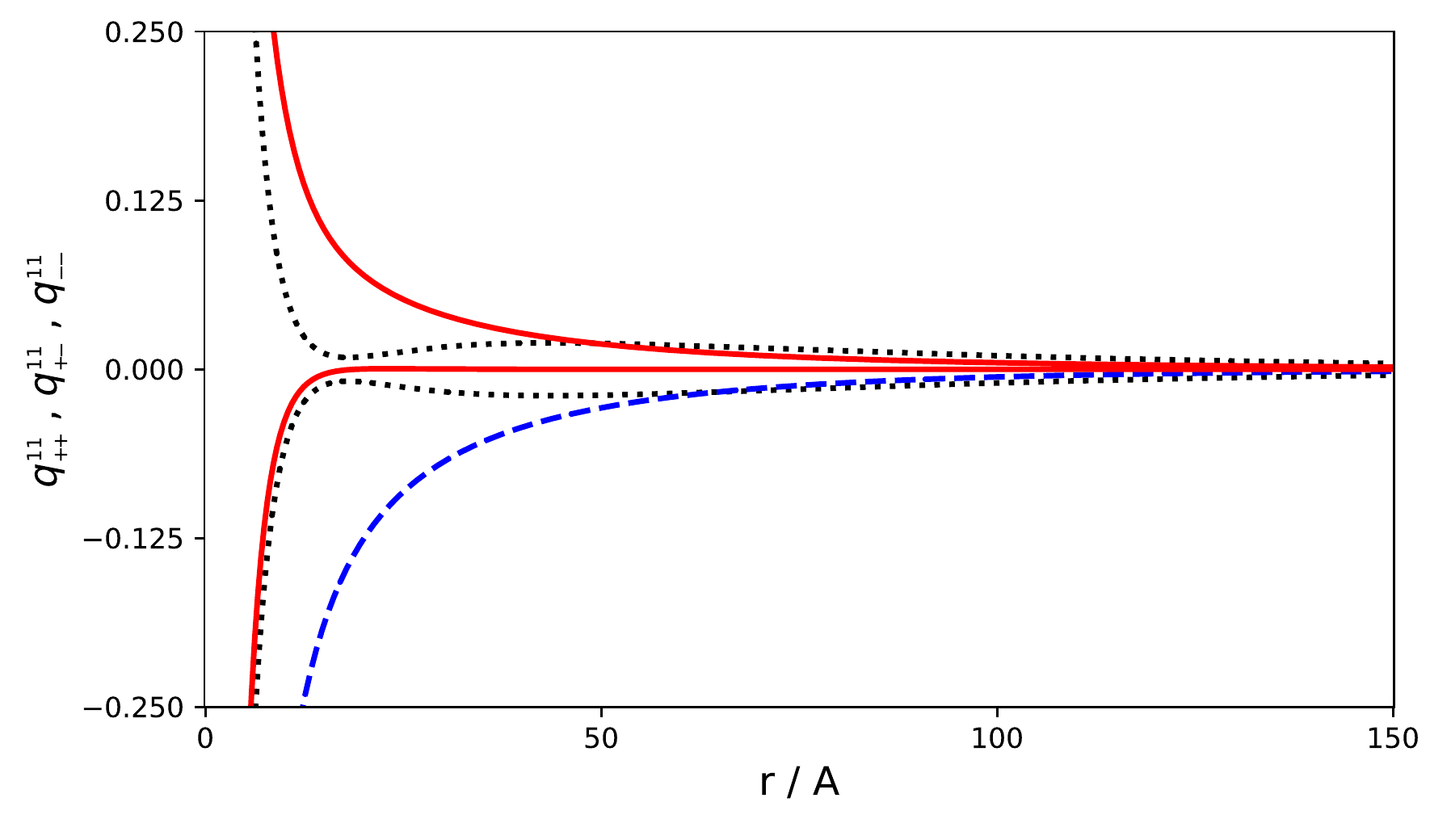}
\includegraphics[scale=0.5, angle=0]{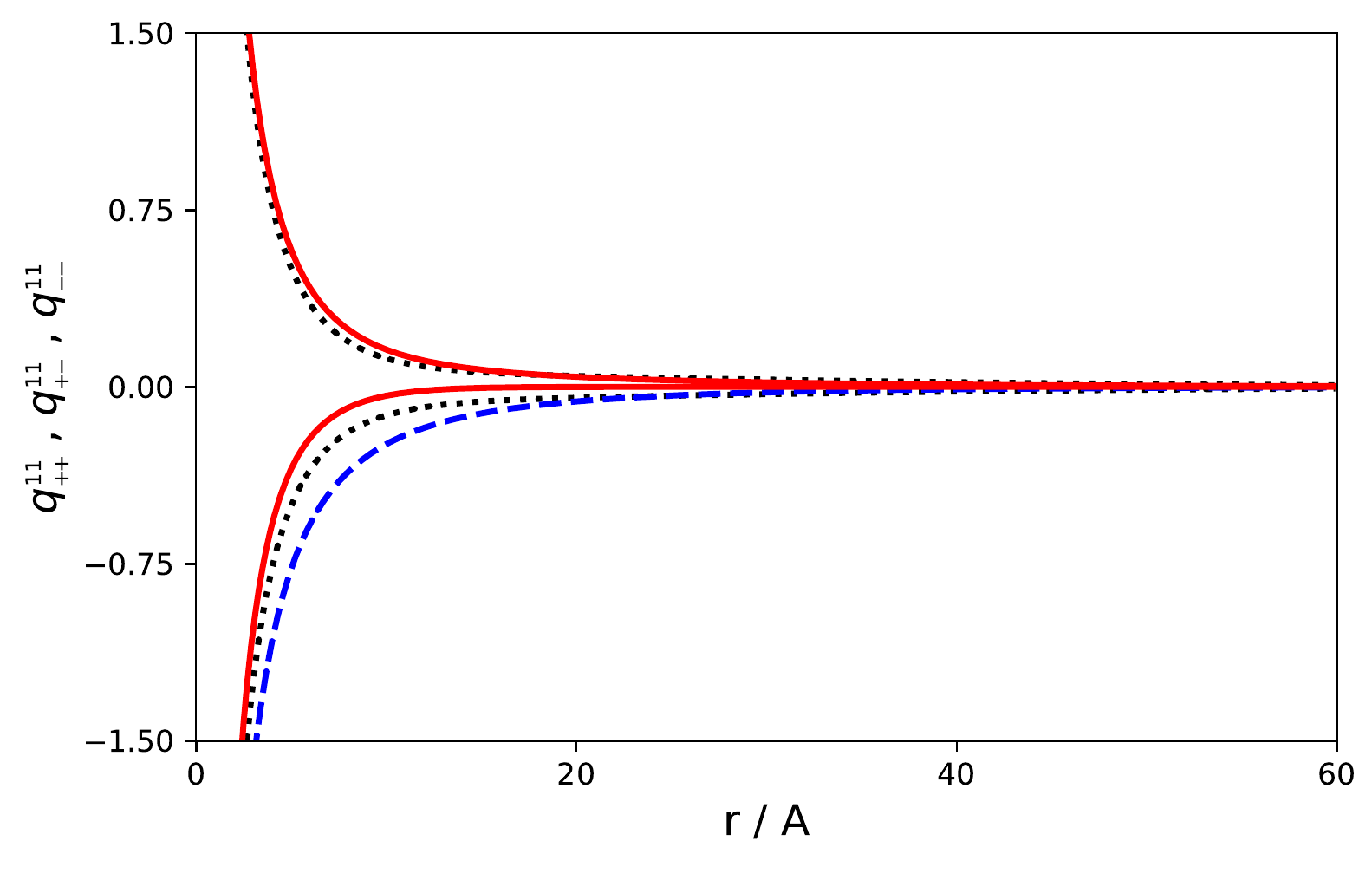}
\caption{(Colour online) ${q}^{11}_{ij}$. Parameters as in figure~\ref{Fig:case1}. The blue dashed lines denote the cation-cation (matrix co-ions) correlations.}
\label{Fig:case2}
\end{figure}

To illustrate how the electro-nonneutrality of matrix and fluid subsystems influences the particle-particle correlations, we plotted ${q}^{10}_{ij}$, and ${q}^{11}_{ij}$ for three sets of parameters (three different fluid concentrations, leading to three different types of screening), that showed unusual qualitative behavior in the case of electroneutral subsytems, and compared them with the corresponding correlations obtained in the electroneutral subsystems~\cite{Hribar1997} containing the same particles. The results for fluid-matrix correlations, and fluid-fluid correlations are shown in figures \ref{Fig:case1}, and \ref{Fig:case2}, respectively. The results of this study are shown with solid red lines (color on-line), and the results for electroneutral subsystems are shown with dotted black lines.

One can see that, as expected, no particular differences are observed for fluid-matrix correlations (${q}^{10}_{ij}$) between the case where the matrix is electroneutral, and in the case of charged matrix (this paper), regardless of the annealed fluid concentration. The fluid-matrix correlations are namely due to the direct interactions between matrix and fluid particles, and these are the same in both cases.

There are, however, differences observed in the case of fluid-fluid correlations (${q}^{11}_{ij}$). In all cases, the interactions between oppositely charged fluid particles, are of similar range as in the case of electroneutral matrix. However, the characteristic sign changing of $--$ and $+-$ function  (crossing of the functions) that is observed at low fluid concentration in the case of electroneutral matrix, does not occur in the case where the matrix carries a net charge. Since it has been established that this crossing occurs as a consequence of matrix-mediated interactions~\cite{Hribar1997}, the result can be explained by the fact that less matrix particles are present in this case. Note that even in the case of electroneutral matrix this kind of behavior is only observed in the cases of low fluid to matrix concentration ratio~\cite{Hribar1997}. The interactions between co-ions of the matrix (dashed blue line in figure~\ref{Fig:case2}), however, are longer ranged (less screening is observed due to lower concentration of this kind of ions). The effect is less pronounced at higher annealed fluid concentrations.

\section{Conclusions}

In this paper we present a rigorous derivation of the charged particle-charged particle interactions in the system, where the matrix and the annealed fluid are both electro-nonneutral. By comparing the results with those, where both subsystems are electroneutral one can see that generally the screening between ions depends on the charge of the subsystems. One should take that into account when developing the renormalization procedure for such systems.

\section{Acknowledgements}
B.H.-L. thanks O. Pizio for helpful discussions on the subject. B.H.-L. also acknowledges the financial support from the Slovenian Research Agency (research core funding No. P1-0201).

\newpage


\ukrainianpart

\title{Екранування іон-іонних кореляцій у розчинах електролітів, адсорбованих у невпорядкованих матрицях: застосування реплічного підходу
	у рівняннях Орнштейна-Церніке}
\author[Т. Млакар, Б. Грібар-Лі]{Т. Млакар,
 Б. Грібар-Лі}
\address{Університет Любляни, факультет хімії та хімічних технологій, вул. Вечна 113, 1000 Любляна, Словенія
}

\makeukrtitle
	\begin{abstract}
		Реплічні рівняння Орнштейна-Церніке для електроліту, адсорбованого у зарядженій невпорядкованій матриці, застосовано для моделі, у якій обидві підсистеми складаються з точкових (додатних або від'ємних) зарядів. Тоді як система в цілому є електронейтральною, кожна з підсистем має певний сумарний заряд. Результати даного дослідження порівнюються з попередніми результатами, коли взаємодії у системі вважались такими самими, як у випадку електронейтральних підсистем.
\keywords	екранування у електролітах, заряджена матриця, реплічний метод Орнштейна-Церніке
		%
	\end{abstract}

\end{document}